\documentclass[twocolumn]{aastex631}
\usepackage{multirow}
\usepackage[tbtags]{amsmath}

\newcommand{\ls}{LS~5039}
\newcommand{\fgl}{4FGL~J1826.2$-$1450}
\newcommand{\gr}{$\gamma$-ray}

\shortauthors{Zeng et al.}
\graphicspath{{./}{figures/}}

\begin{document}

\title{Investigating the Potential of LS 5039 as a Triple System Using Fermi-LAT Data}

\author{Lujun Zeng}
\affiliation{Department of Astronomy, School of Physics and Astronomy, Key Laboratory of Astroparticle Physics of Yunnan Province, Yunnan University, Kunming 650091, People's Republic of China; zhangpengfei@ynu.edu.cn}

\author{Mengqing Zhang}
\affiliation{Department of Astronomy, School of Physics and Astronomy, Key Laboratory of Astroparticle Physics of Yunnan Province, Yunnan University, Kunming 650091, People's Republic of China; zhangpengfei@ynu.edu.cn}

\author{Chongyang Ren}
\affiliation{Key Laboratory of Dark Matter and Space Astronomy, Purple Mountain Observatory, Chinese Academy of Sciences, Nanjing 210034, People’s Republic of China; jzyan@pmo.ac.cn}
\affiliation{School of Astronomy and Space Science, University of Science and Technology of China, Hefei, Anhui 230026, People’s Republic of China}

\author{Pengfei Zhang}
\affiliation{Department of Astronomy, School of Physics and Astronomy, Key Laboratory of Astroparticle Physics of Yunnan Province, Yunnan University, Kunming 650091, People's Republic of China; zhangpengfei@ynu.edu.cn}

\author{Jingzhi Yan}
\affiliation{Key Laboratory of Dark Matter and Space Astronomy, Purple Mountain Observatory, Chinese Academy of Sciences, Nanjing 210034, People’s Republic of China; jzyan@pmo.ac.cn}
\affiliation{School of Astronomy and Space Science, University of Science and Technology of China, Hefei, Anhui 230026, People’s Republic of China}

\begin{abstract}

\ls~is one of a handful of \gr~binary systems in the Milky Way, comprising a pulsar and a massive
O-type companion star with an orbital period of 3.9~day. Recently, we conducted a data
analysis using approximately 16~year of Fermi-LAT observations, spanning from 2008~August~4 to
2024~July~8. In our timing analysis, we discovered two new periodic signals
with frequencies higher and lower than the known orbital period.
The higher-frequency signal has a period of 3.63819~day with a $7.1\sigma$
confidence level, while the lower-frequency signal has a period of 4.21654~day with a $6.3\sigma$
confidence level. Additionally, in data from the High Energy Stereoscopic System of Cherenkov
Telescopes, two potential signals with periods similar to the two newly discovered ones.
Considering that these two signals fall within the same frequency interval as
the orbital period, we suggest the possibility of a third body orbiting the barycenter of the
\ls~binary system, with the new periodic signals arising from specific frequency combinations
of the two orbital periods.

\end{abstract}


\section{Introduction}
\label{Intro}
%Binary systems are composed of a young massive star and a neutron star or a black hole,
%they are the most common type of objects in the Galaxy \citep{liu06,liu07}.
To date, only a few \gr~binary systems have been identified in our
Galaxy \citep{aab+06,alb+06,hin+09,lat+12,lj17,cor+19}, with \ls~being one of them.
\ls~is a well-studied high-mass X-ray binary, consisting of a rapidly rotating neutron
star with a spin period of 9~s \citep{yme+20} and a massive O-type companion
star \citep{gosss3+16} with a mass of 23~$M_\odot$ \citep{cas+05}. The two objects orbit
each other with a period of 3.9~day in a moderately eccentric orbit with an eccentricity
of 0.35 and an orbital inclination of $i=25^\circ$, at a distance of 2.5 kpc from Earth
within our Galaxy \citep{cas+05}. The orbital phase zero (i.e.~$\phi_0$) is defined at
its periastron passage, with the reference epoch $T_0$ set at
HJD 2451943.09 \citep[i.e.~MJD 51942.59;][]{cas+05}.
Notably, \ls~is one of only three known \gr~binary systems in which the compact object
is a neutron star, the other two being LS~I~+~61$^\circ$303 \citep{wen+22} and
PSR B1259$-$63/SS~2883 \citep{jml+92}.

\ls's electromagnetic emissions span a broad spectrum, ranging from radio frequencies to
TeV \gr~energies \citep{mrp12,cas+05,tku+09,ls5039+09,aab+06} and potentially reaching
PeV \citep{cao21,bpp+24}. Its spectral energy distribution, which is dominated by MeV-GeV
\gr s, characterizes LS 5039 as a \gr~binary. The 3.9~day orbital modulation of \ls~has
been identified by multi-wavelength observations and is well-supported by extensive literature.
Notably, at energies above 0.1~TeV, \citet{aab+06} presented its orbital modulation with a
period of $3.9078\pm0.0015$~day based on the dataset from the High Energy Stereoscopic
System (HESS) of Cherenkov Telescopes.
In GeV, \ls~was detected by \citet{ls5039+09} using the 11~month \emph{Fermi}-LAT observations,
who also reported that its \gr~emissions are modulated by a $3.903\pm0.005$~day orbital period.
%It was subsequently cataloged in the first Fermi-LAT source catalog \citep[1FGL;][]{1fgl2010}
%under the designation 1FGL~J1826.2$−$1450.
Recently, \citet{z24} discovered a new periodic signal with a period of 26.3~day
in the \gr~binary LS~I~+~61$^\circ$303 using \emph{Fermi}-LAT observations.
Motivated by these findings, we conducted a new timing analysis of \ls,
hoping to uncover phenomena beyond its known orbital period in GeV gamma-rays.

In the fourth catalog Data Release 4 \citep[4FGL-DR4;][]{4fgl-dr4}, \ls~is associated with
the \gr~counterpart 4FGL~J1826.2$-$1450, which has been monitored by Fermi-LAT for
approximately 16~year. In our timing analysis, in addition to the detection of the orbital
period of 3.90609~day, we identified two new periodic signals
with frequencies higher and lower than the orbital period.
These signals have periods of 3.63819~day and 4.21654~day, with confidence levels of
7.1$\sigma$ and 6.3$\sigma$, respectively.
Moreover, two similar periodic signals were observed in data
from the HESS of Cherenkov Telescopes \citep{aab+06}.
The details of our data analysis and findings are presented below.
%the numbers in the parentheses are their uncertainties on the last digit,
In them, the numbers in parentheses indicate the uncertainties associated with the values.

\section{Data Analysis and Results}
\label{sec:lat-data}

\subsection{Data Reduction}
\label{sec:model}

%%%%%%%%%%%%
\begin{table}
\begin{center}
\caption{Results of likelihood}
\begin{tabular}{ccccccc}
\hline\hline
%\multirow{2}{*}{Parameters} & \multicolumn{3}{c}{PLC} & \multicolumn{3}{c}{LP} \\
%\cline(1-3)
%  &$\Gamma$ & $b$ & $E_c$ & $\alpha$ & $\beta$ & $E_b$ \\
Models & \multicolumn{6}{c}{Parameter values}  \\
%  &$\Gamma$ & $b$ & $E_c$ & $\alpha$ & $\beta$ & $E_b$ \\
\hline
LP & $\alpha$ & $\beta$ & $E_{\rm b}$ & TS & $F_{\rm ph}$ \\
   & 2.758 & 0.102 & 2.0 & & &\\
   & 2.701(3) & 0.036(1) & 2.0 & 29003.2 & 7.46(9)\\
\hline
PLEC & $\Gamma$ & $b$ & $E_{\rm c}$ & TS & $F_{\rm ph}$ \\
   & 2.511(16) & 2/3 & 13.9(31) & 28912.7 & 7.46(17)\\
\hline%\hline
\end{tabular}
\label{tab:par}
\end{center}
{\bf Notes. }{Best-fit parameters of the likelihood, $F_{\rm ph}$ in unit of
              $\times10^{-7}$~photons~cm$^{-2}$~s$^{-1}$, $E_{\rm b}$ and
              $E_{\rm c}$ in unit of GeV.}
\end{table}
%%%%%%%%%%%%

For our data analysis, we utilized the Pass 8 \emph{Front+Back} events (evclass = 128
and evtype = 3), focusing on the energy range of 0.1--500.0~GeV within a $20^\circ\times20^\circ$
region of interest centered on \fgl~(R.A.~=~276.5637, Decl.~=~$-$14.8496), covering the
period from MJD~54682.687 to 60499.299. Events with zenith angles $<90^\circ$ were selected,
and only high-quality events from good time intervals were retained using the expression
“DATA\_QUAL$>$0~\&\&~LAT$\_$CONFIG==1”. The Galactic and isotropic diffuse emissions were
modeled using the templates of gll\_iem\_v07.fits and iso\_P8R3\_SOURCE\_V2\_v1.txt, respectively.
The data was reduced and analyzed using the Fermitools version~2.2.0.

\begin{figure*}
\centering
\includegraphics[angle=0,scale=0.66]{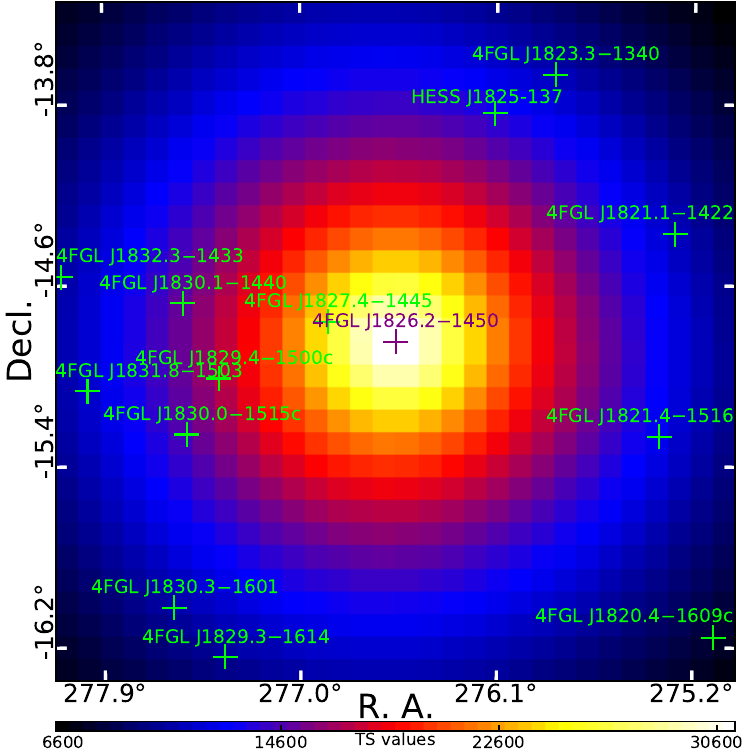}
\includegraphics[angle=0,scale=0.66]{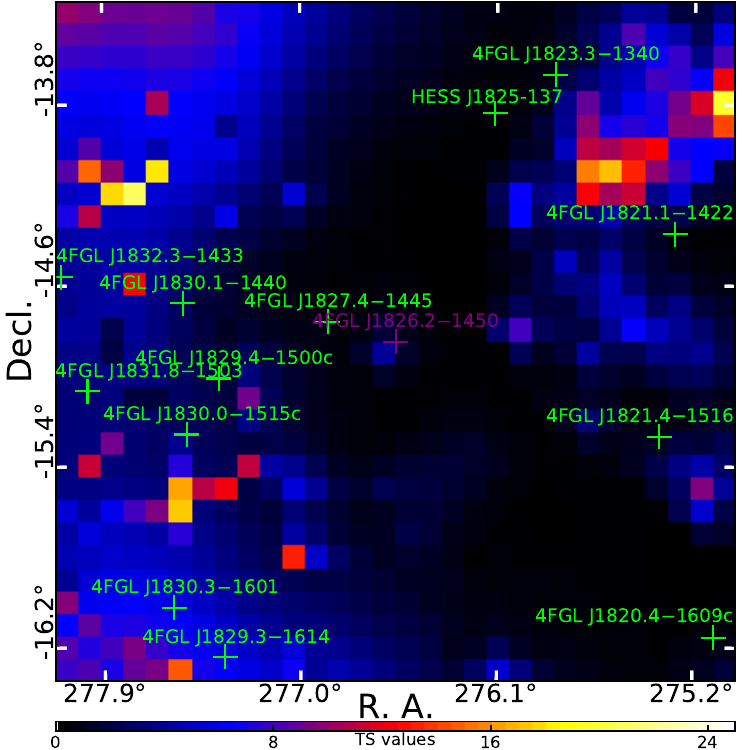}
\caption{TS maps in the 0.1–500.0 GeV range, covering a $3^{\circ}\times3^{\circ}$
         region centered on \fgl, with \gr~sources from the 4FGL-DR4 marked by
         crosses. Left panel: TS map representing \gr~emissions from \fgl, generated
         using the best-fit model with the target excluded.
         Right panel: Residual TS map created using the same model, except with \fgl~included.
         }
\label{fig:tsmap}
\end{figure*}

A model file was created based on 4FGL-DR4, and a binned maximum likelihood analysis was
performed to update the model parameters.
For \fgl, the 4FGL-DR4 provides a log-parabola (LP) spectral shape,
$dN/dE=N_0(E/E_b)^{-[\alpha+\beta\log(E/E_b)]}$.
Additionally, we also employed a power-law with an exponential cutoff (PLEC) model,
$dN/dE=N_0(E/E_0)^{-\Gamma}\exp[-(E/E_{\rm c})^b]$, to fit the events around \fgl.
All parameters for these two models are summarized in Table~\ref{tab:par}.
As shown in it, we conclude that the LP model, which yields
a higher TS value of 29,003.2, provides a better representation of the \gr~emissions from
\ls~compared to the PLEC model, which has a TS value of 28,912.7. The LP model results
were then saved as the best-fit model. Based on it, we generated a TS map for \fgl’s
\gr~emissions (see left panel of Figure~\ref{fig:tsmap}) and a residual TS map excluding
\gr s from all sources listed in the 4FGL-DR4 (right panel of Figure~\ref{fig:tsmap}).
These maps indicate that the \gr~emissions from \fgl~is well-described by the LP model.

%%%%%%%%%%%%

%%%%%%%%%%%%
\begin{figure*}
\centering
\includegraphics[angle=0,scale=0.67]{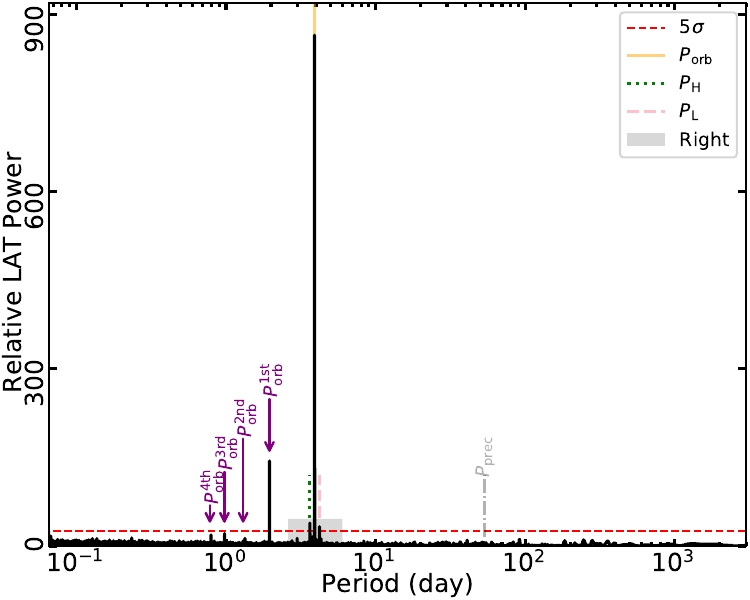}
\includegraphics[angle=0,scale=0.67]{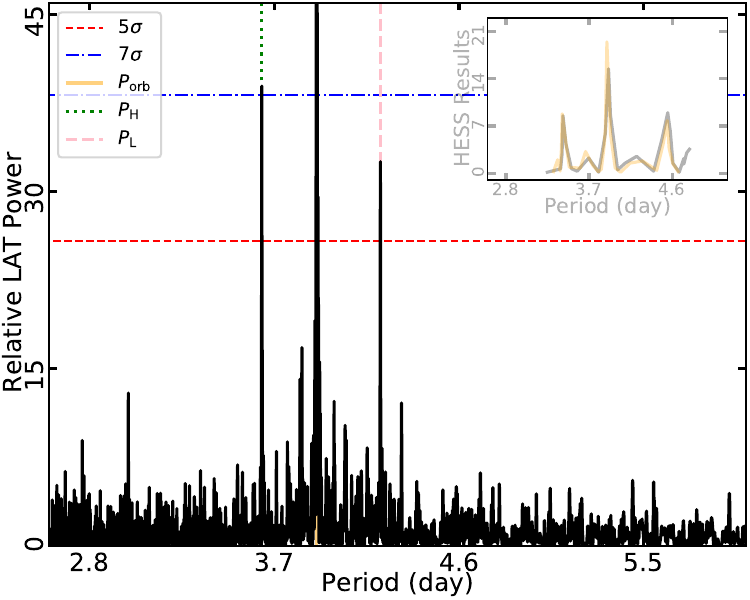}
\caption{LSP power spectra (black histogram) derived from the 0.1--500.0~GeV AP
         light curve of \fgl. Left panel: The three periodicities, $P_{\rm H}$, $P_{\rm orb}$,
         and $P_{\rm L}$, are in close proximity. The gray dashed-dotted line marks
         the location of the 53.4~day precession period of the \emph{Fermi}-LAT orbit.
         Right panel: LSP power spectrum zoomed in the left. The green dotted and pink dashed
         lines indicate the signals of $P_{\rm H}$ and $P_{\rm L}$, respectively.
         The horizontal red dashed and blue dashed-dotted lines stand for $5\sigma$ and $7\sigma$
         confidence levels. HESS results that drawn from the Figures~2~and~3 of \citet{aab+06}
         are shown in the inset plot.}
\label{fig:all-lsp}
\end{figure*}
%%%%%%%%%%%%
\begin{figure}
\centering
\includegraphics[angle=0,scale=0.67]{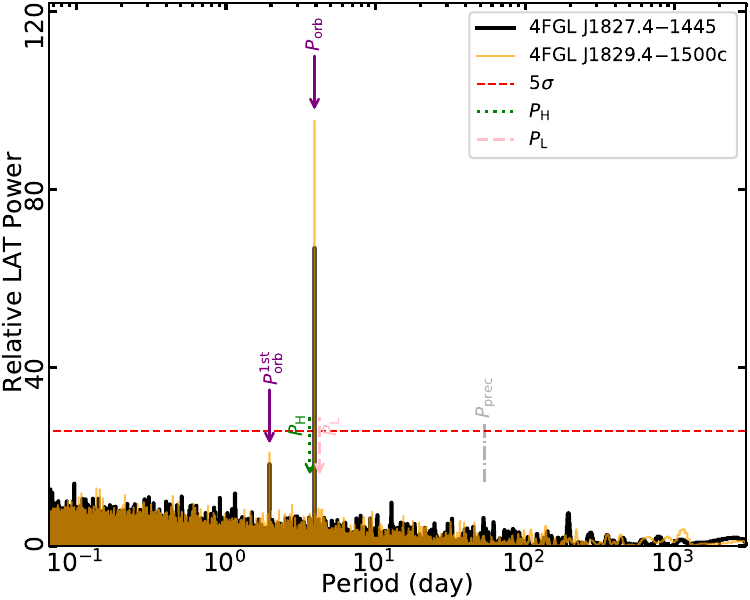}
\caption{LSP power spectra for 4FGL~J1827.4$-$1445 (black histogram) and
         4FGL~J1829.4$-$1500c (orange histogram).
         The vertical green and pink arrows show the positions of the two signals
         of $P_{\rm H}$ and $P_{\rm L}$. The gray dashed-dotted line is
         the precession period of the Fermi-LAT orbit.
         And the red dashed line is 5$\sigma$ confidence level.}
\label{fig:2src-lsp}
\end{figure}
%%%%%%%%%%%%

\subsection{Light Curve Construction and Timing Analysis}
%\subsection{Timing Analysis}
\label{sec:timing}
A modified version of the aperture photometry (AP) method was employed to construct the
light curve for \fgl~in the 0.1--500.0~GeV. To optimize the signal-to-noise ratio,
an aperture radius of $3^\circ.16$ was selected, taking into account the performance
characteristics of the LAT instrument,
with a criterion of $\theta<$ max(6.68$-$1.76log$_{10}(E_{\rm MeV}), 1.3)^{\circ}$, 1.3),
following that performed in \citet{abd+10}. The light curve was binned at 500~s intervals
for each time bin. To mitigate contamination from \gr s originating from the Sun and Moon,
we excluded events occurring when the target was within 5$^{\circ}$ of either the Sun or the
Moon, using \emph{gtmktime}. To address significant exposure time variations across different
time bins, the exposure time for each bin was calculated using \emph{gtexposure}. Using the
best-fit model, event probabilities for \fgl~were determined by employed \emph{gtsrcprob} and
used as weights for constructing the AP light curve, rather than simply counting photon
numbers \citep{k11,J1018+12,cor+19}. Additionally, the times in the light curve were corrected
to the barycenter using \emph{gtbary}.

In our timing analysis, a method of the generalised Lomb–Scargle
periodogram \citep[LSP;][]{l76,s82,zk09} was employed to derive the power spectrum for
the AP light curve.
Compare to other period search algorithms, the LSP offers more accurate frequency determination
for unevenly spaced data, is less susceptible to aliasing, provides better frequency resolution,
improves the detection of weak signals, and facilitates the assessment of the confidence level
for periodic signals. This same methodology was also employed by \citet{z24}.
The power spectrum covers a frequency range from $f_{\rm max}$~=~1/0.06
day$^{-1}$ to $f_{\rm min}$~=~1/5816.612~day$^{-1}$
(e.g., the inverse of the length of \emph{Fermi}-LAT observations, $1/T_{\rm obs}$).
We show the power spectrum in left panel of Figure~\ref{fig:all-lsp},
in it a prominent main peak appears at 3.90609(3)~day,
%1/3.906092366283404171 day = 0.2560103311001544 +/- 2.190953232078032e-06
corresponding to the orbital period ($P_{\rm orb}$) of LS 5039. In addition to $P_{\rm orb}$,
two peaks can be observed on either side of it. One period appears at the higher frequency
with period of 3.63819(14)~day ($P_{\rm H}$) and another at the lower frequency at
4.21654(20)~day ($P_{\rm L}$).
%1/3.638186651047901066 day = 0.27486220359584124+/-1.032927938984663e-05;
%1/4.216538812949367276~day = 0.2371613411760638+/-1.120632479546412e-05
%0.27486220359584124 - 0.2560103311001544 = 0.018851872495686817
%0.2560103311001544 - 0.2371613411760638 = 0.01884898992409062
For the three periods, their uncertainties were calculated
by $\delta P=\frac{3}{8}\frac{P^2}{T_{\rm obs}\sqrt{p_{_{\rm n}}}}$, where $P$ is a period,
$T_{\rm obs}$ is the length of \emph{Fermi}-LAT observations,
and $p_{_{\rm n}}$ is the normalized height of the peak power,
as that provided in \citet{hb86}.
These periods, $P_{\rm orb}$, $P_{\rm H}$, and $P_{\rm L}$, correspond to frequencies of
$f_{\rm orb}=0.256010(2)$~day$^{-1}$, $f_{\rm H}=0.274862(10)$~day$^{-1}$, and
$f_{\rm L}=0.237161(11)$~day$^{-1}$, respectively. Moreover, the frequency difference between
$P_{\rm H}$ and $P_{\rm orb}$ ($f_{\rm H}-f_{\rm orb}=0.018851$) is nearly equal to the
difference between $P_{\rm orb}$ and $P_{\rm L}$ ($f_{\rm orb}-f_{\rm L}=0.018849$).
In left panel of Figure~\ref{fig:all-lsp}, we marked the $P_{\rm orb}$, $P_{\rm H}$, and $P_{\rm L}$ with orange
solid, green dotted, and pink dashed lines, respectively. And the first, second, third, and fourth harmonics
of $P_{\rm orb}$ also appear, we mark them with four purple arrows.
For clarity we zoom in the power spectrum in the gray shaded region in right panel of Figure~\ref{fig:all-lsp}.

The heights of the peaks of $P_{\rm orb}$, $P_{\rm H}$ and $P_{\rm L}$
are $\sim865.88$, $\sim36.38$, and $\sim33.10$, respectively, compared to the mean power level,
based on the normalization provided in \citet{hb86}.
For $P_{\rm H}$ and $P_{\rm L}$, the probabilities ($p_{_{\rm lsp}}$) of a periodic signal
reaching a power level by chance fluctuation assuming the Gaussian white noise were calculated
based on method provided by \citet{l76} and \citet{s82}. And their values are $1.20\times10^{-17}$
and $4.21\times10^{-15}$, respectively. In our timing analysis, the number of independent frequencies
(i.e., the trial factor) was calculated by $N=(f_{\rm max}-f_{\rm min})/\delta f$ = 96943,
where $\delta f$ is frequency resolution determined by the length of the \emph{Fermi}-LAT observations.
Then the False Alarm Probability (FAP) is calculated by the form of
FAP=$1-(1-p_{_{\rm lsp}})^N\sim p_{_{\rm lsp}}\times N$, after taking into account the trial
number $N$. And their FAPs are derived to be $1.16\times10^{-12}$ and $4.09\times10^{-10}$
for $P_{\rm H}$ and $P_{\rm L}$, which corresponding to $\sim7.1\sigma$ and $\sim6.3\sigma$
confidence levels. In right panel of Figure~\ref{fig:all-lsp}, we show $5\sigma$ and $7\sigma$ confidence levels
with red dashed and blue dashed-dotted lines, respectively.
Interestingly, the HESS data reported by \citet{aab+06} also present two potential signals
with periods of 3.42(5) and 4.56(10)~day, which are similar to the two signals reported here.
The errors associated with these periods were derived from the full-widths at half-maximum of
the power peaks. For easy reference, we also include a schematic of the results for \ls~in the
inset of right panel of Figure~\ref{fig:all-lsp}, adapted from \citet{aab+06}.

Considering the broad point spread function of the \emph{Fermi}-LAT, periodic signals from
neighboring sources may also appear in the power spectrum of the target. From the residual TS map
(right panel of Figure~\ref{fig:tsmap}), we can see that there are no new \gr~sources beyond those listed
in the 4FGL-DR4. Therefore, we constructed power spectra for the two \gr~sources closest to the target,
4FGL~J1827.4$-$1445 and 4FGL~J1829.4$-$1500, using the same process,
and present them in Figure~\ref{fig:2src-lsp} with black and orange histograms, respectively.
We observe a prominent peak corresponding to the $P_{\rm orb}$ of \ls, along with a weaker peak
corresponding to its first harmonic. However, no signals corresponding to $P_{\rm H}$ and
$P_{\rm L}$ are detected.

%%%%%%%%%%%%
\begin{figure}
\centering
\includegraphics[angle=0,scale=0.68]{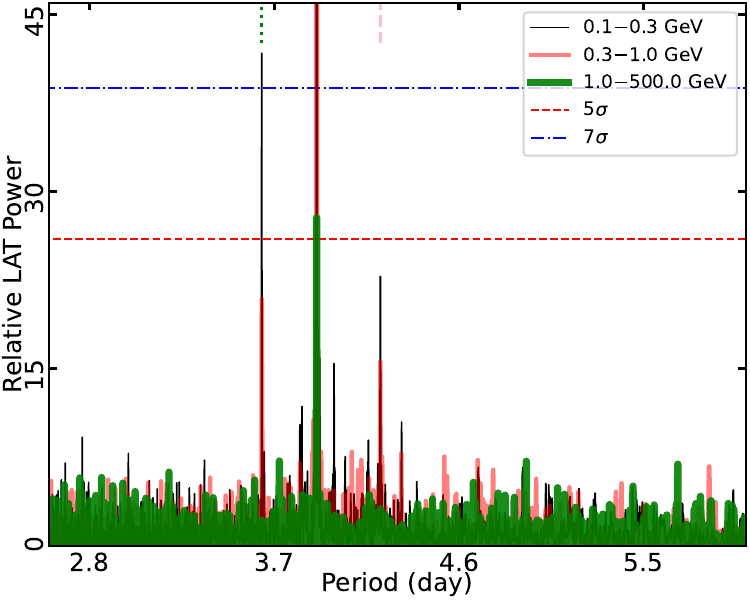}
\caption{\ls's energy-dependent LSP power spectra in 0.1--0.3, 0.3--1.0, and 1.0-500.0~GeV.
         The black, red, and green histograms stand for the spectra constructed with 0.1--0.3,
         0.3--1.0, and 1.0--500.0~GeV AP light curves, respectively.
         As in Figure~\ref{fig:all-lsp}, the red dashed and blue dashed-dotted lines
         are 5$\sigma$ and 7$\sigma$ confidence levels.}
\label{fig:ende}
\end{figure}
%%%%%%%%%%%%

To test whether these two signals are energy-dependent, we constructed light curves in three energy
bands, 0.1--0.3, 0.3--1.0, and 1.0--500.0~GeV, using the same aperture photometry method.
Their power spectra are shown in Figure~\ref{fig:ende} with black, red, and green
histograms, respectively. We observe that the peak values of these two new signals, as well as the
orbital period signal, decrease as the energy increases across the three energy ranges.

%%%%%%%%%%%%
\begin{figure}
\centering
\includegraphics[angle=0,scale=0.86]{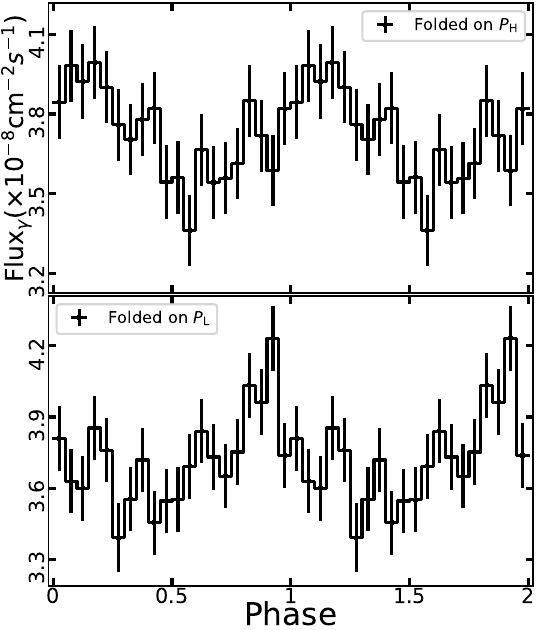}
\caption{Phase-resolved light curves derived from the 0.1--500.0~GeV Fermi-LAT events,
         folded over the periods $P_{\rm H}$ (top panel) and $P_{\rm L}$ (bottom panel).
         Both panels share the same y-axis labels.}
\label{fig:phlc}
\end{figure}
%%%%%%%%%%%%

\subsection{Phase-resolved Analysis}
\label{sec:phlc}
To examine the periodic profiles of the two signals, we derived phase-resolved light curves
based on the periods of $P_{\rm H}$ and $P_{\rm L}$ by dividing the 0.1–500.0 GeV Fermi-LAT
events into 20 phase intervals. For each phase interval, we performed a likelihood analysis
to determine the fluxes, using phase zero at periastron and setting the reference epoch $T_0$
at MJD 51,942.59, as described by \citet{crr+05}. We present the phase-resolved light curves
for $P_{\rm H}$ and $P_{\rm L}$ in the upper and lower panels of Figure~\ref{fig:phlc},
respectively. These light curves clearly demonstrate periodic modulation profiles that differ
from the modulation profile of the orbital period. Moreover, there are also significant
differences between the profiles of $P_{\rm H}$ and $P_{\rm L}$.

\section{Summary and Discussion}
\label{sec:dis}
%\ls~is one of the rare \gr~binaries in the Milky Way, consisting of a rapidly rotating
%neutron star \citep{yme+20} and a massive O-type companion star \citep{crr+05}.
Here, we conducted a timing analysis using the LSP method on approximately 16 years of
\emph{Fermi}-LAT observations. Our analysis identified two new periodic signals
with frequencies higher and lower than the orbital period (3.90609~day).
These signals have periods of 3.63819~day and 4.21654~day, with confidence levels
of 7.1$\sigma$ and 6.3$\sigma$, respectively. Moreover, these new
periodic signals share the same frequency interval as the orbital period.
All three signals, including the two new periodic signals and the orbital period,
show a trend of decreasing strength with increasing energy. This trend may be attributed
to statistical factors, as there are much more \gr s in lower energies than that in higher energies.
Additionally, in previous studies,
the HESS data appeared two potential signals with similar periods \citep{aab+06}
to those we discovered in our Fermi-LAT analysis, which independently support
the authenticity of our new findings.
We know that in astrophysics, there are many causes that can produce periodic signals,
but it is quite rare for two signals to have the same frequency interval.
Considering that the $P_{\rm H}$ and $P_{\rm L}$ signals share the same interval as the
orbital period in the frequency domain, we will discuss two scenarios that could generate
these signals in the following discussion.

During timing analysis, peaks with the same frequency interval in the power spectrum 
can originate from the sidelobes
of the main peak \citep[i.e., the orbital period;][]{b76}, caused by the finite length
of observations and the use of windowing functions.
These sidelobes are evenly distributed on both sides of the main peak,
with their peak values gradually decreasing,
which makes the first sidelobe the most likely candidates.
In this scenario, we can obtain the locations of the first sidelobe using the formula:
$P_{\rm sidelobe}=P_{\rm orb}\pm\frac{7}{5}P_{\rm orb}^{2}/T_{\rm obs}$ in our timing analysis,
where $P_{\rm sidelobe}$ is the periods corresponding to the first sidelobe,
and $T_{\rm obs}$ is the length
of Fermi-LAT observations during the timing analysis. The values for the first sidelobes
are calculated to be 3.902420 or 3.909765~day, which differ significantly from the
%3.9024200259439845 or 3.909764706622824~day
$P_{\rm H}$ and $P_{\rm L}$ values. Therefore, we believe that the $P_{\rm H}$ and
$P_{\rm L}$ signals cannot originate from the sidelobes of the orbital period of \ls.
Additionally, if the two signals originated from the sidelobes,
they should also appear in the power spectra of the two neighboring \gr~sources. However,
the power spectra of these two sources are significantly influenced by the orbital
period of \ls, and they do not show the $P_{\rm H}$ and $P_{\rm L}$ signals. This further
indicates that the $P_{\rm H}$ and $P_{\rm L}$ signals are related to \ls~and not
associated with the sidelobes.

We then introduce an alternative scenario that could also produce two periodic signals
with the same frequency interval.
We propose that there may be a third body in the LS 5039 system, suggesting that
\ls~could be a three-body system, as that reported in \citet{3bd+14,3bd+22}.
The third body orbits the barycenter of the original binary system,
with a hypothesized orbital period of $P_{\rm hpy}$.
And $P_{\rm H}$ and $P_{\rm L}$ signals likely originate
from certain frequency combinations involving the $P_{\rm orb}$ and $P_{\rm hyp}$.
To simplify this process, we omit the phase shift parameters and assume that the $P_{\rm hyp}$
modulation has a sinusoidal function. And the \gr~flux ($F_{\gamma}$) is modulated by a form of
\begin{equation}
F_{\gamma}=A\sin(2\pi t f_{\rm hyp}) + C.
\label{eq:f}
\end{equation}
Since the $P_{\rm orb}$ signal is much stronger than the $P_{\rm H}$ and $P_{\rm L}$
in the power spectrum (Figure~\ref{fig:all-lsp}), hence we assume the amplitude ($A$)
and constant ($C$) of $P_{\rm hyp}$ are strongly modulated by the main periodic signal ($P_{\rm orb}$) and
they have the forms of
\begin{equation}
A=A_1\sin(2\pi t f_{\rm orb}) + C_1
\label{fm:a}
\end{equation}
and
\begin{equation}
C=A_2\sin(2\pi t f_{\rm orb}) + C_2.
\label{fm:c}
\end{equation}
Then we simplify this process again and assume $A_1=A_2=1$ and $C_1=C_2=0$. Then,
%\begin{equation}
\begin{align}
F_{\gamma}= 0.5\{\cos[2\pi t (f_{\rm orb} &- f_{\rm hyp})] - \cos[2\pi t (f_{\rm orb} + f_{\rm hyp})]\} \notag\\& + \sin(2\pi t f_{\rm orb}).
\label{fm:fm}
\end{align}
%\end{equation}
%%%%%% old one
%\begin{equation}
%F_{\gamma}= 0.5\{\cos[2\pi t (f_{\rm orb} - f_{\rm hyp})] - \cos[2\pi t (f_{\rm orb} + f_{\rm hyp})]\} + \sin(2\pi t f_{\rm orb}).
%\label{fm:fm}
%\end{equation}
%%%%%%
Consequently, we will see two peaks (with the corresponding frequencies to
$f_{\rm H}=f_{\rm orb}+f_{\rm hyp}$ and $f_{\rm L}=f_{\rm orb}-f_{\rm hyp}$)
%of $P_{\rm H} = P_{\rm orb}P_{\rm hyp}/(P_{\rm orb}-P_{\rm hyp})$
%and $P_{\rm L} = P_{\rm orb}P_{\rm hyp}/(P_{\rm orb} + P_{\rm hyp})$
around the orbital period in the power spectrum. The hypothesized period can be calculated as
$P_{\rm hyp}=\frac{1}{f_{\rm hyp}}=\frac{2}{f_{\rm H}-f_{\rm L}}=53.049(21)$~day, with the
uncertainty derived from the propagation of errors in $f_{\rm H}$ and $f_{\rm L}$.

Additionally, we note that \emph{Fermi}-LAT operations can introduce artificial periodic
signals \citep{fermi09}.
One such signal, which is close to the $P_{\rm hyp}$, corresponds to the \emph{Fermi}-LAT's orbital
precession period of
53.4~day\footnote{https://fermi.gsfc.nasa.gov/ssc/data/analysis/LAT\_caveats\_temporal.html},
we indicate it in Figures~\ref{fig:all-lsp} (left) and \ref{fig:2src-lsp}
by gray dashed-dotted lines.
However, there are no peaks at this precession period in the power spectra of \fgl~and
the two neighboring \gr~sources, and neither $P_{\rm H}$ or $P_{\rm L}$ appears in the power
spectra of the neighboring sources.
Therefore, we conclude that our light curves are not influenced by
this precession period, furthermore the 53.4~day period does not fall within the uncertainties of $P_{\rm hyp}$.
Moreover, the HESS dataset also reveal two potential signals \citep{aab+06} with
periods similar to those reported here.
Thus, we infer that $P_{\rm H}$ or $P_{\rm L}$ are not related to the \emph{Fermi}-LAT's orbital
precession period.

Regardless, this discovery enhances our knowledge of LS 5039 and offers new insights
into the dynamics and evolution of massive binary systems.
The cause of these two new periodic signals remains uncertain.
Further observations, particularly multi-wavelength ones, are encouraged to confirm the
authenticity of the two periodic signals and to reveal their origins.

\begin{acknowledgments}
We would like to thank the anonymous referee for helpful suggestions and Jian Li for useful comments.
This work is supported in part by the National Natural Science Foundation of 
China under grant Grants Nos.~12233006, U2031205, and 12163006, the Basic Research Program of Yunnan Province 
No.~202201AT070137, and the joint foundation of Department of Science and Technology
of Yunnan Province and Yunnan University No.~202201BF070001-020.
P.Z. acknowledges the support by the Xingdian Talent Support Plan - Youth Project.
\end{acknowledgments}

\bibliographystyle{aasjournal}
\bibliography{aas}
\end{document}